\documentstyle[12pt]{article}
\renewcommand{\baselinestretch}{1}
\begin{document}
\title{Dispersive contribution to CP violation in hyperon decays}
\author{P. \.{Z}enczykowski\\
\\
Department of Theoretical Physics\\
The H. Niewodnicza\'nski Institute of Nuclear Physics\\
Radzikowskiego 152, Krak\'ow, Poland\\
}
\maketitle
\begin{abstract}
Within the standard model,
CP-violating asymmetries $A(\Lambda ^o_-)$ and $A(\Xi ^-_-)$ have
been estimated in short-distance calculations to
occur at the level of a few times $10^{-5}$.  We show here that
in this model the dispersive contribution tends
to give asymmetries of similar size.
\end{abstract}

\newpage
\baselineskip = 10 pt
\renewcommand{\baselinestretch}{1.685}
\small
\normalsize

A lot of interest is now directed towards understanding the origin of CP
violation in particle physics \cite{WW}. Among various processes
considered as promising areas of experimental study, the nonleptonic
hyperon decays offer the possibility to observe
CP violation in $\Delta S = 1$ transitions. The upcoming E871
experiment is expected to reach sensitivity of $10^{-4}$ for the sum of
CP-violating asymmetries $A(\Lambda ^0_-)$ and $A(\Xi ^-_-)$ \cite{Russ}.
Estimates of these asymmetries within standard model
yield values at a few times $10^{-5}$ level
\cite{DHP86HSV91,HV95}.
It has been stressed \cite{DHP86HSV91} that good understanding of the
dominant CP-conserving part of nonleptonic hyperon decays as well as
knowledge of strong interaction effects (and, in particular, of scattering
phases) are indispensable if we are to draw meaningful conclusions from
any observed signal of CP violation.

Recently an economic and successful model of the CP-conserving
part of hyperon nonleptonic decays has been proposed \cite{Zen94}.
The success of the model stems from allowing a substantial
(relative to $W$-exchange) total penguin contribution in
baryon-to-baryon weak transitions (see Fig.1).
In agreement with general requirements of chiral and weakly broken
$SU(3)$ symmetries the contributions from diagrams with weak
Hamiltonian acting in the meson leg were assumed non-dominant.
Without introducing {\em any} additional parameters
the model of ref.\cite{Zen94} generates
a relation between the values of the
$f/d$ ratios for S- and P- waves,
which agrees with experiment very well.
Thus, it is plausible that the origin of the difference in the values
of the $f/d$ ratios for the two waves has been identified.
As a by-product the approach fixes then the size of (the real part of)
the total penguin contributions in the S- and P- waves.

Short-distance estimates of the CP-conserving part of penguin contribution
in hyperon decays yield values that fall short of the data by a
factor of at least 5
\cite{DGHP0}.
A similar factor of 5 is needed in kaon decays as well.
Arguments have been presented that such factors presumably originate
from long-range hadron-level effects
\cite{Cheng,Zen95,Zen92}.
If hadron-level effects are indeed that important, then for
a reliable estimate of the CP-violating penguin contribution
one has to consider both hadron-level (dispersive) and short-distance
effects. However, the estimates of CP violation in hyperon decays
carried out so far are based on short-distance calculations only
\cite{DHP86HSV91,GilWise,KaonCPVlong}.

In order to estimate the dispersive effects
we assume in the following that
the total penguin contribution to nonleptonic hyperon decays
is dominated by hadron-loop effects.
If these effects are smaller the numbers calculated below
should be scaled down appropriately.

The data fix the reduced $SU(6)$
matrix elements corresponding to $W$-exchange and penguin diagrams
($(b1)+(b2)$ and
$(c1)+(c2)$ of Fig.1) to be (in units of $10^{-7}$, see  ref.\cite{Zen94})

(1) for S-waves:
\begin{eqnarray}
\label{eq:Sbc}
b_S & = & -5  \nonumber \\
c_S & = & 12
\end{eqnarray}
with $(f/d)_S = -1 + 2c_S/(3b_S) = -2.6$

2) for P-waves:
\begin{eqnarray}
\label{eq:Pbc}
b_P & = & -132 \nonumber \\
c_P & = &  158
\end{eqnarray}
with $(f/d)_P = -1 + 2c_P/(3b_P) = - 1.8$.

The obtained description of $\Lambda ^0_-$ and $\Xi ^-_-$ decays is
compared with experiment in Table 1 (for a more complete account see
ref.\cite{Zen94}). It is known that the
$F/D$ ratio describing the SU(3) structure of strong couplings relevant
in the expressions for the P-wave amplitudes is around 0.57 which is
slightly different from its $SU(6)$ value of 2/3 implicitly assumed in
Eq.(\ref{eq:Pbc}).
Description of P-wave amplitudes for $F/D=0.57$ is given in Table 1
as well.

Table 1.
Full amplitudes and penguin contributions in CP-conserving amplitudes of
nonleptonic hyperon decays.

\begin{center}
\begin{tabular}{|c|l|c|c|c|c|}
\hline
               & wave & SU(3) expression & penguin & total  & experiment
\\
\hline
$\Lambda ^0_-$ &  S   &
$-\frac{1}{2\sqrt{6}}b_S + \frac{1}{2\sqrt{6}}c_S$
& 2.45 & 3.47 & 3.23 \\
\hline
         & P~~~~(F/D) &                  &       &          &       \\
         & ~~~~~~~~$2/3$&
$\frac{1}{6\sqrt{6}}b_P + \frac{1}{2\sqrt{6}}c_P$
& 32.3  & 23.3     &  22.1 \\
         & ~~~~~~~$0.57$&                  & 29.2  & 17.6     &       \\
\hline
$\Xi ^-_-$     &  S   &
$\frac{1}{\sqrt{6}}b_S - \frac{1}{2\sqrt{6}}c_S$
& -2.45 & -4.49 & -4.50 \\
\hline
         & P~~~~(F/D) &                  &       &          &       \\
         & ~~~~~~~~$2/3$&
$-\frac{1}{3\sqrt{6}}b_P - \frac{1}{6\sqrt{6}} c_P$
&  -10.8  & 7.2    &  16.6 \\
        & ~~~~~~~$0.57$ &                  &  -7.7 & 15.5     &       \\
\hline
\end{tabular}
\end{center}

The hadron-level penguin contribution arises from
that part of hadronic loops which - when viewed at the quark level -
generates quark loops with $q$ $=$
$u$,$c$, and $t$ (Fig.2). Its real part - of interest to us - may be
estimated through dispersion relations.
Contribution from hadronic states involving $t$ quarks may be safely
neglected: the relevant thresholds are very remote. On the other hand,
dispersive effects which correspond to $u$ and $c$ quark loops
have to be considered.
Although we will neglect direct contribution from the top states, the
existence of $t$ quark is felt in the $u$, $c$ sector as a deviation
from equality of
$V^*_{ud}\,V^{}_{us}$ and $-V^*_{cd}\,V^{}_{cs}$:
\begin{equation}
\label{eq:orthogon}
V^*_{cd}\,V^{}_{cs} = - V^*_{ud}\,V^{}_{us}-
V^*_{td}\,V^{}_{ts}
\end{equation}
Thus, the sum of contributions from intermediate
hadron states corresponding to quark loops is
\begin{eqnarray}
\label{eq:CPviol}
&V^*_{ud}\,V^{}_{us}\,L_H(u)+V^*_{cd}\,V^{}_{cs}\,L_H(c)
+V^*_{td}\,V^{}_{ts}\,L_H(t)  \approx & \nonumber \\ & \approx
V^*_{ud}\,V^{}_{us}\,(L_H(u)-L_H(c)) - V^*_{td}\,V^{}_{ts}\,L_H(c) &
\end{eqnarray}
where dependence on KM factors has been explicitly factored out of
hadron-level-induced loop contribution $L_H(q)$ of quark q.
The second term on the r.h.s. of Eq.(\ref{eq:CPviol})
induces CP violation.
The size of the CP-violating term
is governed by
\begin{equation}
\label{eq:V/C}
{\rm Im}\, \tau \equiv {\rm Im}\,\frac{- V^*_{td}V_{ts}}{V^*_{ud}V_{us}}
= A^2 \lambda ^4 \eta \le 10^{-3}
\end{equation}
(using $\lambda = 0.22 $, $A = 0.9\pm 0.1$, and $\eta \le 0.4 \pm 0.2$
    ).
Large mass of the $c$ quark produces an additional suppression factor
$s_c$, so that in fact
the scale of CP violating amplitudes relative to the CP-conserving one
is set by
\begin{equation}
\label{eq:sc}
{\rm Im}\, \tau \, \, s_c \equiv {\rm Im}\, \tau \, \,
L_H(c)/(L_H(u)-L_H(c))
\end{equation}
To estimate the ratio of dispersive factors let us note that the
long-range
penguin contribution originates from weak transitions that occur
within hadrons from
intermediate meson+baryon virtual states. Loop factors $L_H(q)$ are
proportional to the probability of finding such a virtual state in
a physical baryon and -
with $SU(4)$ symmetric couplings -
they are inversely proportional to the squares of energy denominators.
Thus,
\begin{equation}
\label{eq:ratio}
L_H(c)/L_H(u) = [(E_{M_u}+E_{B_u}-E)/(E_{M_c}+E_{B_c}-E)]^2
\end{equation}
where $E$ is total baryon energy, and
$E_{M_q}$, $E_{B_q}$ are energies of the intermediate meson and
baryon containing quark $q$.
In a dispersive calculation
the dominant contributions to dispersive integrals arise
when intermediate meson (baryon) momenta $k$ are in the range of
$k^2 \approx 0.6~-~1.2~GeV^2 $.
Indeed, the numerator in the integrand of the dispersion relation
has a form of roughly
\begin{equation}
\label{eq:spectral}
k^3 \exp (-(k/k_{cutoff})^2)
\end{equation}
where the first factor comes from phase-space and the p-wave
character of strong virtual decays, and the second factor
arises from hadronic formfactors and provides the cutoff.
In realistically sized hadrons $k_{cutoff} \approx 0.7\,GeV$
\cite{TorZen},
whence the above range of $k^2$.
One estimates then that
$L_H(c)/L_H(u) \approx 0.1-0.2 $.
In conclusion, dispersive calculations give
$s_c \approx 10-20$\%.
Calculation of CP asymmetries is now straightforward.
Normalizing the penguin contribution to the full amplitude according to
Table 1
one obtains the following weak phases
\begin{eqnarray}
\label{eq:weakangles}
\phi _S(\Lambda ^0_-) & = & + 0.7\, s_c\, {\rm Im}\, \tau \nonumber \\
\phi _P(\Lambda ^0_-) & \approx & + 1.35\, s_c\, {\rm Im}\, \tau \nonumber \\
\phi _S(\Xi ^-_-)     & = & + 0.54\, s_c\, {\rm Im}\, \tau \nonumber \\
\phi _P(\Xi ^-_-)     & \approx & -0.55\, s_c\, {\rm Im}\, \tau
\end{eqnarray}
The CP-violating asymmetries are then calculated from \cite{DHP86HSV91}
\begin{equation}
\label{eq:CPVasym}
A = - \tan(\delta _P - \delta _S) \sin (\phi _P -
\phi _S)
\end{equation}
where $\delta _{P,S}$ are phase shifts due to final-state strong interactions.
With Im $\tau  \approx 10^{-3}$ and
using $\delta _P (\Lambda ^0_-) = -1.1 ^o$,
$\delta _S (\Lambda ^0_-) = +6.0 ^o $,
$\delta _P (\Xi ^-_-) = -2.7 ^o $, and
$\delta _S (\Xi ^-_-) = -18.7 ^o $
from ref.\cite{RoperNath}
one gets
\begin{eqnarray}
\label{eq:final}
A(\Lambda ^0_-) & = & 0.081\, s_c\, {\rm Im}\, \tau \approx (0.8-1.6) \cdot
10^{-5}
\nonumber \\
A(\Xi ^-_-)     & = & 0.31\, s_c\, {\rm Im}\, \tau \approx (3-6) \cdot 10 ^{-5}
\end{eqnarray}
Thus, dispersive effects tend to give
\begin{equation}
\label{eq:Asum}
A(\Lambda ^0_-)+A(\Xi ^-_-) = (3.8-7.6) \cdot 10 ^{-5}
\end{equation}
i.e. a number at the same "a few times $10^{-5}$ level" as the
short-distance estimates.
If for the $\Xi ^-_-$ decays one uses
smaller strong rescattering phases
(as calculated recently)
the $A(\Xi ^-_-)$ asymmetry is scaled down accordingly.

This research has been supported in part by the Polish Committee for
Scientific Research Grant No. 2 P03B 231 08.

\newpage
\small
\normalsize
\renewcommand{\baselinestretch}{1.685}
\setlength{\unitlength}{0.58pt}
\begin{picture}(550,450)
\put(89,80){
\begin{picture}(460,360)
\put(0,190){
\begin{picture}(200,160)
\put(100,15){\makebox(0,0){(b1)}}
\put(170,90){\vector(-1,0){25}}
\put(145,90){\line(-1,0){50}}
\multiput(130,65)(0,5){5}{\line(0,1){3}}
\put(65,90){\vector(-1,0){20}}
\put(45,90){\line(-1,0){15}}
\put(65,150){\vector(0,-1){30}}
\put(65,120){\line(0,-1){30}}
\put(95,90){\vector(0,1){30}}
\put(95,120){\line(0,1){30}}
\put(170,65){\vector(-1,0){90}}
\put(80,65){\line(-1,0){50}}
\put(170,40){\vector(-1,0){90}}
\put(80,40){\line(-1,0){50}}
\end{picture}}
\put(0,10){
\begin{picture}(200,160)
\put(100,15){\makebox(0,0){(c1)}}
\put(170,90){\vector(-1,0){15}}
\put(155,90){\line(-1,0){10}}
\put(115,90){\line(-1,0){20}}
\multiput(115,90)(5,0){6}{\line(1,0){3}}
\put(130,90){\oval(30,30)[b]}
\put(130,75){\vector(-1,0){0}}
\put(65,90){\vector(-1,0){20}}
\put(45,90){\line(-1,0){15}}
\put(65,150){\vector(0,-1){30}}
\put(65,120){\line(0,-1){30}}
\put(95,90){\vector(0,1){30}}
\put(95,120){\line(0,1){30}}
\put(170,65){\vector(-1,0){90}}
\put(80,65){\line(-1,0){50}}
\put(170,40){\vector(-1,0){90}}
\put(80,40){\line(-1,0){50}}
\end{picture}}
\put(250,190){
\begin{picture}(200,160)
\put(100,15){\makebox(0,0){(b2)}}
\put(170,90){\vector(-1,0){15}}
\put(155,90){\line(-1,0){20}}
\put(135,90){\vector(0,1){30}}
\put(135,120){\line(0,1){30}}
\put(105,150){\vector(0,-1){30}}
\put(105,120){\line(0,-1){30}}
\put(170,65){\vector(-1,0){50}}
\put(120,65){\line(-1,0){90}}
\put(170,40){\vector(-1,0){50}}
\put(120,40){\line(-1,0){90}}
\put(105,90){\vector(-1,0){50}}
\put(55,90){\line(-1,0){25}}
\multiput(70,65)(0,5){5}{\line(0,1){3}}
\end{picture}}
\put(250,10){
\begin{picture}(200,160)
\put(100,15){\makebox(0,0){(c2)}}
\put(170,90){\vector(-1,0){15}}
\put(155,90){\line(-1,0){20}}
\put(135,90){\vector(0,1){30}}
\put(135,120){\line(0,1){30}}
\put(105,150){\vector(0,-1){30}}
\put(105,120){\line(0,-1){30}}
\put(170,65){\vector(-1,0){50}}
\put(120,65){\line(-1,0){90}}
\put(170,40){\vector(-1,0){50}}
\put(120,40){\line(-1,0){90}}
\put(105,90){\line(-1,0){20}}
\multiput(55,90)(5,0){6}{\line(1,0){3}}

\put(55,90){\vector(-1,0){10}}
\put(45,90){\line(-1,0){15}}
\put(70,90){\oval(30,30)[b]}
\put(70,75){\vector(-1,0){0}}
\end{picture}}

\end{picture}}
\end{picture}
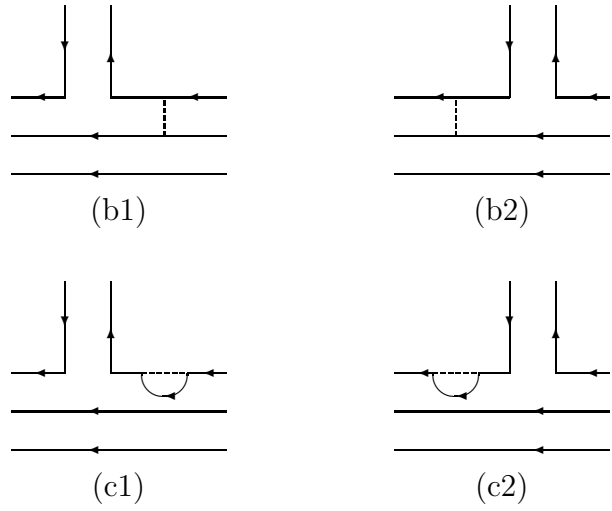
\begin{center} {Fig.1. Quark diagrams for weak decays}
\end{center}
\vfill
\newpage

\setlength {\unitlength}{1.6pt}
\begin{picture}(260,260)
\put(30,160){\begin{picture}(150,100)
\put(85,50){\line(1,0){45}}
\put(10,50){\line(1,0){45}}
\put(70,50){\oval(30,30)}
\put(10,55){$B_f$}
\put(120,55){$B_i$}
\put(45,65){$B$}
\put(86,65){$B'$}
\put(65,24){$M$}
\put(70,65){\circle*{5}}
\put(66,75){$H_{W}^{p.c.}$}
\end{picture}}
\put(15,10){
\begin{picture}(170,100)
\put(0,70){\line(1,0){40}}
\put(40,80){\oval(20,20)[rb]}
\put(60,80){\oval(20,20)[lt]}
\put(60,90){\line(1,0){50}}
\put(110,80){\oval(20,20)[rt]}
\put(130,80){\oval(20,20)[lb]}
\put(170,70){\vector(-1,0){40}}
\put(0,60){\line(1,0){85}}
\thicklines
\put(100,60){\vector(-1,0){15}}
\put(70,70){\oval(20,20)[lt]}
\put(70,80){\line(1,0){15}}
\thinlines
\put(85,80){\line(1,0){15}}
\put(100,70){\oval(20,20)[rt]}
\put(120,70){\oval(20,20)[lb]}
\put(120,60){\line(1,0){10}}
\put(170,60){\vector(-1,0){40}}
\thicklines
\put(65,37){$q=u,c,t$}
\put(60,70){\line(0,-1){30}}
\put(70,40){\oval(20,20)[lb]}
\put(70,30){\vector(1,0){30}}
\put(100,40){\oval(20,20)[rb]}
\put(110,40){\line(0,1){10}}
\put(100,50){\oval(20,20)[tr]}
\thinlines
\put(0,50){\line(1,0){40}}
\put(40,40){\oval(20,20)[tr]}
\put(50,40){\line(0,-1){10}}
\put(60,30){\oval(20,20)[bl]}
\put(60,20){\line(1,0){50}}
\put(110,30){\oval(20,20)[rb]}
\put(120,30){\line(0,1){10}}
\put(130,40){\oval(20,20)[lt]}
\put(170,50){\vector(-1,0){40}}
\multiput(85,61)(0,4){5}{\line(0,1){2}}
\put(90,68){$W$}
\end{picture}
}
\put(100,150){\vector(0,-1){30}}
\end{picture}
\\
\\
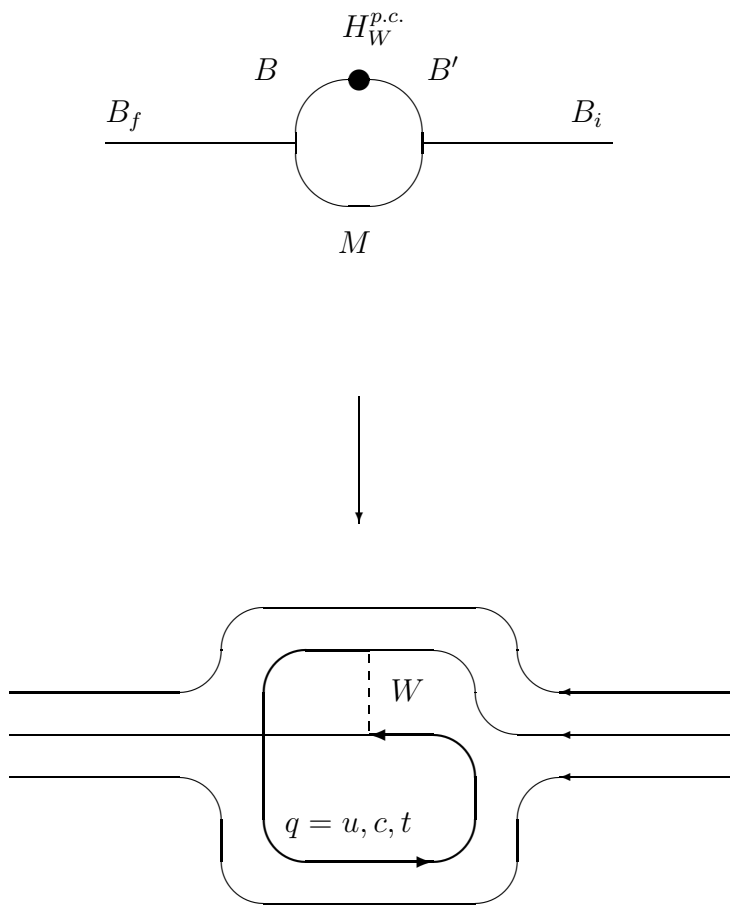
Fig.2. Example of quark-loop generation from a hadron-level loop
diagram for the baryon-to-baryon
matrix element of the parity conserving part of the weak Hamiltonian.
\end{document}